\documentstyle[12pt]{article}

\catcode`@=11 \@addtoreset{equation}{section} \catcode`@=12

\newcommand{\affa}{Theory Group, Tata Institute of Fundamental Research,
   Homi Bhabha Road, Bombay 400005, India, Email--- sgupta@theory.tifr.res.in}
\newcommand{\affb}{Lehrstuhl f\"ur Theoretische Physik IV,
   D-44221, Dortmund, Germany, Email--- indu@hal1.physik.uni-dortmund.de}
\newcommand{\affc}{Centre for Theoretical Studies, Indian Institute of
   Science, Bangalore 560012, India, Email--- prakash@cts.iisc.ernet.in}
\newcommand{\affd}{Theory Group, Tata Institute of Fundamental Research,
   Homi Bhabha Road, Bombay 400005, India, Email--- ravi@theory.tifr.res.in}
\begin{document}
\begin{titlepage}
\begin{flushright}\vbox{\begin{tabular}{c}
           TIFR/TH/95-05\\
           DO-TH-95-03\\
           IISC-CTS-1/95\\
           May 25, 1995\\
           hep-ph/9507341
\end{tabular}}\end{flushright}
\begin{center}
   {\large \bf  Bloch-Nordsieck Thermometers:\\
                One-loop Exponentiation in\\
                Finite Temperature QED.}
\end{center}
\bigskip
\begin{center}
   {Sourendu Gupta\footnote\affa,
    D.\ Indumathi\footnote\affb,
    P.\ Mathews\footnote\affc,
    V.\ Ravindran\footnote\affd.}
\end{center}
\bigskip
\begin{abstract}
We study the scattering of hard external particles in a heat bath in
a real-time formalism for finite temperature QED. We investigate the
distribution of the 4-momentum difference of initial and final hard
particles in a fully covariant manner when the scale of the process,
$Q$, is much larger than the temperature, $T$. Our computations are
valid for all $T$ subject to this constraint. We exponentiate the
leading infra-red term at one-loop order through a resummation of
soft (thermal) photon emissions and absorptions. For $T>0$, we find
that tensor structures arise which are not present at $T=0$. These
carry thermal signatures. As a result, external particles can serve
as thermometers introduced into the heat bath. We investigate the
phase space origin of $\log(Q/m)$ and $\log(Q/T)$ terms.
\end{abstract}
\end{titlepage}
\setcounter{footnote}{0}

\section{Introduction\label{intro}}

Recently there has been a lot of phenomenological interest
in external particles interacting inside a heat-bath. This
is due to the realisation that hard jets formed in the early
stages of a heavy-ion collision can interact with a quark-gluon
plasma formed later, and perhaps carry away information on
this plasma \cite{gyulassy,gupta}. However, such processes are
often infra-red sensitive. Phenomenological studies
have used arbitrary infra-red cutoffs to obtain finite results.
It is legitimate to ask whether one can remove this arbitrariness
and obtain stable finite results for such infra-red sensitive processes.

In this paper we consider the simpler, but related, problem of hard
scatterings of some external particles introduced within a heat-bath
at a temperature $T$.
We choose two of the simplest processes in QED--- the scattering
of an electron with an off-shell photon and the annihilation of an
electron-positron pair into an off-shell photon.
Due to quantum and thermal fluctuations, extra soft particles can
also be involved in each of these scatterings.

If $p^{\rm in}$ and $p^{\rm out}$ are the summed momenta of all
the hard initial and final state particles respectively, then one
can construct the difference
\begin{equation}
  P_\mu\;=\;p^{\rm in}_\mu-p^{\rm out}_\mu.
\label{in:diff}\end{equation}
We study the distribution of $P$ through the measurement of a suitable
semi-inclusive cross section. It is non-trivial because of the extra
soft particles. When $P^2\ll Q^2$ ($Q^2$ is the scale of the hard
process), the Lorentz-invariant cross section factorises as
\begin{equation}
  {d^4\sigma\over dP^4}\;\approx\;
                   \sigma_0{d^4{\cal P}\over dP^4},
\label{in:factor}\end{equation}
where $\sigma_0$ is the inclusive cross section for the process. Throughout
this paper we assume such a factorisation, and study the distribution
of $P_\mu$ (the second factor on the right hand side) through its cumulants,
{\sl i.e.\/}, the connected parts of the moments.

Naively, it might seem that the heat-bath spoils the Lorentz
invariance of various cross sections. This is incorrect. The
bath only selects a frame; Lorentz invariance cannot be lost
due to the existence of a special frame. Cross sections which
are Lorentz invariant at $T=0$ remain so for $T>0$. We carry
out a fully covariant treatment of the resummation and show
that the difference between $T=0$ and $T>0$ physics lies in
some new asymmetries and angular correlations. These are
proportional to powers of the temperature. As a result,
scatterings of external particles can be used as thermometers
for the heat-bath. This is the main result of this paper.

It is useful to recall the status of infra-red sensitive
cross sections at $T=0$. Exclusive scattering cross sections can be
infra-red singular. The analysis due to Bloch and Nordsieck \cite{bn}
showed that such measurements are impossible to make with any real
detector. They argued that a real detector, with a finite energy
resolution, measures, at best, some semi-inclusive cross section.
They computed one such cross section and showed that it is indeed
non-singular. In the limit where the detector becomes ideal,
{\sl i.e.\/}, the resolution becomes infinitely sharp, this cross
section remains non-singular.

A stronger theorem is also available for $T=0$ field theories; it
goes by the name of the Kinoshita-Lee-Nauenberg theorem \cite{kln}.
The theorem states that inclusive cross sections are infra-red
finite, order by order in perturbation theory. Infra-red singularities
in real and virtual diagrams cancel in each order. Semi-inclusive
cross sections are also infra-red finite. However, incomplete
cancellations between real and virtual parts show up as large
logarithms. It is possible to sum these to all orders in
perturbation theory.

At finite temperatures the infra-red singularities could be stronger,
since every squared matrix element comes with a thermal distribution
function attached to each external leg. Each external boson then adds
one power to the infra-red divergence. A $T>0$ version of the KLN
theorem has not been proved. Computations have been performed in
several models to check whether or not infra-red divergences cancel
in suitably defined inclusive or semi-inclusive processes
\cite{old,altherretal}.

An important recent result is due to Landshoff and Taylor \cite{lt}.
In a leading order perturbative computation of the decay width for
a scalar going to two scalars, they found exact cancellation of the
infra-red singularity. Weldon introduced a crucial simplification
in finite temperature QED \cite{wel94}. By eikonalising the soft
part of the one-loop order cross section, he was able to resum
soft photon effects to all orders in perturbation theory and obtain
the distribution $d{\cal P}/dP_0$. This is an extension of the
Bloch-Nordsieck theorem to finite temperature.

This proof is a crucial input to our work. We extend it first to the
case of interest to us, {\sl i.e.\/}, the full momentum resummation.
We construct a Lorentz invariant characteristic function for the
distribution of $P$, by exponentiating the first order perturbative
correction to the hard process. Then we extract the cumulants of this
distribution and find terms of the form $\log(Q/m)$. These large
logarithms come from parts of phase space where the extra photons
are nearly collinear with the fermion momenta, exactly as at $T=0$.
These large terms allow us to identify phenomenologically interesting
results from such soft photon resummations at finite temperature.
We also find terms in $\log(Q/T)$. These arise from terms left over
after cancellation of the infra-red divergence.

The plan of this paper is the following. In Section \ref{s:resum}
we shall perform the resummation and show that Weldon's proof of
infra-red finiteness \cite{wel94} can be generalised in a Lorentz
invariant way to yield the distribution of $P$. An outline of a
diagrammatic proof of the cancellation of infra-red divergences is
given in Appendix \ref{ap:resum}. In Section \ref{zerot}
we examine the mean and covariance of $P_\mu$ at $T=0$, since this
is the base against which the $T>0$ results have to be compared.
In Section \ref{ft} the corresponding $T>0$ quantities are computed.
Various details of these calculations are relegated to Appendices
\ref{ap:kinema}--\ref{ap:tensor}. Our results and their implications
are discussed in Section \ref{results}. The reader who is not
interested in the details of the computation may skip to this section.

\section{The Resummed Distribution\label{s:resum}}

In this paper we assume that a factorisation of the hard and soft
parts of the cross section holds in the form given in
eq.~(\ref{in:factor}), and then examine the soft part of the cross
section, {\sl i.e.\/}, the distribution of $P_\mu$. The generating
function for all matrix elements needed for a resummation of the
soft part of the one-loop order contribution is \cite{wel94},
\begin{equation}
  Z(J,T)\;=\;\int{\cal D} A_\mu\exp\left[-i\int d^4x\left(
   {1\over4}F_{\mu\nu}F^{\mu\nu}+J_\mu A^\mu\right)\right].
\label{model}\end{equation}
The integral over $x_0$ in the exponent is carried out over a path
in the complex time plane. The choice of this path is made so that
the thermal averaging of the S-matrix elements is done properly.
Detailed discussions of the path of integration can be found in
\cite{realtime}. Some more information can be found in Appendix
\ref{ap:resum}.

At this point it is useful to introduce some notation. For the
scattering process, the initial and final momenta of the electron
are $p$ and $p'$, and the initial photon has spacelike 4-momentum
$q$ ($q^2=-Q^2$). For the annihilation process, the electron momentum
is $p$, the positron momentum is $p'$ and the final photon momentum
$q'$ is timelike ($q'^2=Q^2$). Details of the hard scattering
kinematics are given in Appendix \ref{ap:kinema}.

A proof of the factorisation implicit in eq.~(\ref{in:factor}) can be
obtained by considering the leading perturbative corrections to the
hard scattering process and taking the limit of the extra photon
momentum, $k$, going to zero (see Appendix \ref{ap:resum}). For the
processes we consider, this gives
\begin{equation}
  J_\mu(k)\;=\;ie\left({p'_\mu\over p'\cdot k}
      -{p_\mu\over p\cdot k}\right).
\label{current}\end{equation}
Here $e$ is the positron charge. It is important to remember that
this approximation is valid only in the infra-red. Hence, in
order to use eqs.~(\ref{model}) and (\ref{current}) consistently,
we must truncate the mode expansion of the soft fields in the
partition function.

We perform this truncation in a Lorentz invariant fashion by writing
the mode expansion---
\begin{equation}\begin{array}{rl}
  A_\mu(x)\;=\;&\displaystyle\sum_\lambda\displaystyle\int\tilde{dk}
   \left[\epsilon_\mu^*(k,\lambda)a_\lambda^\dagger(k)
           {\rm e}^{ik\cdot x}+\epsilon_\mu(k,\lambda)a_\lambda(k)
           {\rm e}^{-ik\cdot x}\right],\\
   \tilde{dk}\;=\;&
    {\displaystyle d^4k\over\displaystyle(2\pi)^4}2\pi\delta(k^2)
      \Theta(k_0)\Theta(\eta V^2-V\cdot k).
\end{array}\label{mode}\end{equation}
Here $\epsilon_\mu(k,\lambda)$ is the polarisation vector for a
photon of momentum $k$ and polarisation $\lambda$, and $a_\lambda(k)$
and $a_\lambda^\dagger(k)$ are the annihilation and creation
operators. $V$ is a vector built out of the momenta of the hard particles
which sets the scale of the hard process and $\eta\ll1$. The second
theta-function in the Lorentz invariant phase space measure expresses
the fact that the fields are soft.

Note that such a factorised approach is useful only when there is a
great difference between the hard and soft scales. In particular,
we take all thermal fluctuations into the soft scale, and hence we
must have $Q^2\gg T^2$. For both the processes, we choose $V$ to be
the timelike vector $q'$ ($q'=p+p'$ in the hard-scattering kinematics).
The arguments in this section do not depend on the choice of $V$.

In the Feynman gauge, the generating function involves
only Gaussian integrals. Performing them we obtain
\begin{equation}
  Z(J,T)\;=\;\exp\left(-{i\over2}\int{d^4k\over(2\pi)^4}
   J^\mu(k) G_{\mu\nu}(k,u) J^\nu(-k)\right),
\label{partition}\end{equation}
where the thermal photon propagator is given by
\begin{equation}
  G_{\mu\nu}(k,u)\;=\;-g_{\mu\nu}\left\{{1\over k^2+i\epsilon}
 -2\pi i\delta\left(k^2\right)B(k\cdot u;T)\right\},
\label{prop}\end{equation}
and the Bose distribution function is
\begin{equation}
  B(k\cdot u;T)\;=\;{1\over{\rm e}^{|k\cdot u|/T}-1}.
\label{bosefn}\end{equation}
The Lorentz vector $u$ ($u^2=1$) specifies the 4-velocity of the
heat bath with respect to the observer \cite{wel82}. In the rest frame
of the heat bath, $k\cdot u=k_0$, and the Bose distribution takes its
familiar form. This is a covariant way of expressing the fact that the
presence of the heat bath selects out a preferred frame.

The matrix element for $n$-photon emission is obtained by taking
appropriate derivatives of $Z(J,T)$ with respect to
the external current, giving
\begin{equation}
  {\cal M}(k_1,\cdots,k_n)\;=\; Z(J,T)\prod_{i=1}^n
    \epsilon_\mu(k_i) J^\mu(k_i).
\label{matel}\end{equation}
Emission of photons corresponds to taking $k_i^0>0$, and absorption
to $k_i^0<0$. The squared matrix element, after summation over
polarisations, gives a factor of $J^2$ for each photon. Finite
temperature effects in the cross-section may be completely taken
into account through an extension of the usual trick (see Appendix
\ref{ap:resum}) of writing a phase space element
\begin{equation}
  d\phi_i\;=\;{d^4k_i\over(2\pi)^4} 2\pi\delta\left(k_i^2\right)
 \left[\Theta\left(k_i^0\right)+B(k_i\cdot u;T)\right]
 \Theta(\eta q'^2-q'\cdot k_i).
\label{dlips}\end{equation}

The resummation now gives the probability density for radiation as
\begin{equation}
  {d{\cal P}\over d^4P}\;=\;\left|Z(J,T)\right|^2
    \sum_{n=0}^\infty {1\over n!} \int \prod_{i=0}^n
      \left\{d\phi_i J_\mu(k_i) J^\mu(k_i)\right\}
    \delta^4\left(P^\mu-\sum_{i=0}^nk_i^\mu\right).
\label{resum}\end{equation}
For further analysis we construct the Lorentz invariant characteristic
function of this distribution---
\begin{equation}
  \Xi(x)\;\equiv\;\int d^4P {\rm e}^{iP\cdot x}
   {d{\cal P}\over d^4P}\;=\;
        \exp\bigl[R_0(x)+R_{\scriptscriptstyle T}(x)\bigr].
\label{charfn}\end{equation}
The Fourier transformation decouples the delta function, allowing
an exponentiation in the form shown.

The logarithm of the characteristic function is defined by the
expressions
\begin{equation}\begin{array}{rl}
  R_0\;&=\;\displaystyle\int\tilde{dk}
      J(k)\cdot J(k)\left\{{\rm e}^{ik\cdot x}-1\right\},\\
  R_{\scriptscriptstyle T}\;&=\;\displaystyle\int\tilde{dk}
      J(k)\cdot J(k) B(k\cdot u;T)\left\{{\rm e}^{ik\cdot x}
                              +{\rm e}^{-ik\cdot x}-2\right\}.
\end{array}\label{expon}\end{equation}
The subtractions come from $|Z(J,T)|^2$ and may be obtained from
eq~(\ref{partition}). All infrared singularities cancel between the
exponential terms and the constant piece in both the $T=0$ part,
$R_0$ and the $T>0$ contribution, $R_{\scriptscriptstyle T}$. This
is an extension of Weldon's proof of infrared finiteness in this
model \cite{wel94}.

The characteristic function, $\Xi$, is a generating function for the
moments of $P_\mu$. The logarithm, $R_0+R_{\scriptscriptstyle T}$, is
a generating function for the ``connected parts'' of these moments,
{\sl i.e.\/}, the cumulants. The $n$-th cumulant of $P_\mu$ is given by
\begin{equation}
  \left\langle P_{\mu_1}P_{\mu_2}\cdots P_{\mu_n}\right\rangle_c\;=\;
        (-i)^n\left.
    {\partial\over\partial x_{\mu_1}}{\partial\over\partial x_{\mu_2}}
     \cdots{\partial\over\partial x_{\mu_n}}
            \bigr(R_0+R_{\scriptscriptstyle T}\bigr)\right|_{x=0},
\label{cumdef}\end{equation}
where the subscript $c$ denotes a connected part.
Each cumulant of $P$ then becomes the sum of a $T=0$ and a $T>0$
part.

Since the integral representation of $R_{\scriptscriptstyle T}$
is even in $x$, odd derivatives vanish. As a result, there is no
$T>0$ contribution to the odd cumulants. The physics behind this is
simple. At finite temperature the charged particle can remove momentum
from the heat bath by absorbing a photon, as well as add momentum by
emitting. In thermal equilibrium these two processes balance out,
giving a distribution even in $P$. It is possible to pursue this
further. The symmetry under $x\leftrightarrow-x$ comes from the
phase space in eq.~(\ref{dlips}), in particular from the $\Theta+B$
factor there. This is obtained by summing the weights for absorption,
$B(k_0)\Theta(-k_0)$, and emission, $[1+B(k_0)]\Theta(k_0)$. These
two weights are related through the commutation relations of the
photon field operators. In the ultimate analysis, this is the
origin of the vanishing of odd cumulants in thermal equilibrium.

It is useful to note the following result. Combining eqs.~(\ref{expon})
and (\ref{cumdef}), we find that a cumulant of the resummed
distribution is equal to the corresponding moment of the distribution
obtained from the soft part of the leading order in perturbation theory.

We write down explicitly the first cumulant, {\sl i.e.\/}, the mean
radiated momentum---
\begin{equation}
  \left\langle P_\mu\right\rangle\;=\;-i\left.
    {\partial\over\partial x_\mu}\bigl(R_0+R_{\scriptscriptstyle T}
        \bigr)\right|_{x=0}.
\label{re:mean}\end{equation}
$\left\langle P_\mu\right\rangle$ is given entirely by the $T=0$ piece,
\begin{equation}
  \left\langle P_\mu\right\rangle\;=\;\int\tilde{dk}
     J(k)\cdot J(k) k_\mu.
\label{re:meanz}\end{equation}
This is a Lorentz vector. We discuss it in detail later.

We also write down the second cumulant $\langle P_\mu P_\nu\rangle_c=
\langle P_\mu P_\nu\rangle-\langle P_\mu\rangle\langle P_\nu\rangle$.
This is the covariance of the radiated 4-momentum---
\begin{equation}\begin{array}{rl}
  \left\langle{P_\mu P_\nu}\right\rangle_c\;&=\;-\left.
    {\displaystyle\partial\over\displaystyle\partial x_\mu}
        {\displaystyle\partial\over\displaystyle\partial x_\nu}
            \bigl(R_0+R_{\scriptscriptstyle T}\bigr)
                                \right|_{x=0}\;=\;V^0_{\mu\nu}
      +V^{\scriptscriptstyle T}_{\mu\nu},\quad{\rm where}\\
  V^0_{\mu\nu}\;&=\;\displaystyle\int\tilde{dk}
      J(k)\cdot J(k) k_\mu k_\nu,\\
  V^{\scriptscriptstyle T}_{\mu\nu}\;&=\;\displaystyle\int\tilde{dk}
           J(k)\cdot J(k) 2B(k\cdot u;T)k_\mu k_\nu.
\end{array}\label{covar}\end{equation}
The covariance is a symmetric Lorentz tensor of rank 2. The diagonal
components of this tensor give the variance of the corresponding
component of the radiated momentum. The off-diagonal elements specify
the correlation between different momentum components. In particular,
the space-space components quantify angular correlations in the
radiated momentum.

\section{Zero Temperature\label{zerot}}

The mean radiated momentum after resummation can be written in terms
of all the vectors available in the problem, multiplied by appropriate
coefficient functions (see Appendix \ref{ap:tensor}). As a result---
\begin{equation}
   \langle P_\mu\rangle\;=\;\int\tilde{dk}J_\lambda(k)J^\lambda(k)k_\mu
       \;=\;F_1\left({p_\mu\over m}\right)
           +F_2\left({p'_\mu\over m}\right).
\label{mean:decompose}\end{equation}
It is important to note that the frame dependence of all the tensors
is taken entirely into the basis tensors. As a result, the coefficient
functions are Lorentz invariant. One can obtain these and the tensor
components quite trivially by contracting $\left\langle P_\mu\right\rangle$
with each of the vectors on the right hand side.

For convenience, these Lorentz invariant contractions are evaluated in
the Breit Frame (BF), {\sl i.e.\/}, the frame in which $p$ and $p'$ are
equal and oppositely directed,
\begin{equation} \begin{array}{rll}
  I_1\;\equiv\;&{\displaystyle p_\mu\over\displaystyle m}\langle P_\mu\rangle
     &\;=\;{\displaystyle \alpha\over\displaystyle \pi}\eta Q
             \left[(2\Psi-1)\Omega_1-\Omega_6\right],\\
  I_2\;\equiv\;&{\displaystyle p'_\mu\over\displaystyle m}\langle P_\mu\rangle
     &\;=\;{\displaystyle \alpha\over\displaystyle \pi}\eta Q
             \left[(2\Psi-1)\Omega_1-\Omega_6\right].
\end{array} \label{mean:integrals}\end{equation}
Here $\alpha=e^2/4\pi$ and
$\Psi$ is defined in eq.~(\ref{ap:invariants}) and the integrals
$\Omega_i$ in eq.~(\ref{ap:angular}).
The final expressions are obtained in the BF and we have used the fact
that $\Omega_i=\Omega'_i$ in this frame (see Table \ref{ap:angtab} in
Appendix \ref{ap:tensor}). The
equality of the two contractions translates into the equality
of $F_1$ and $F_2$. In the hard-scattering limit ($Q\to\infty$),
we find
\begin{equation}
  F_1\;=\;F_2\;=\;{\alpha\over\pi}{\eta Q\over\bar\gamma}
           \left[2\log\left({Q\over m}\right)-1\right],\\
\label{mean:forms}\end{equation}
where $\bar\gamma$ is defined in eq.~(\ref{ap:bf}).
The logarithm of $Q/m$ arises from that part of the
phase space in which emitted or absorbed photons are collinear
with the electron momenta.

The components of the mean radiated momentum can now be easily
evaluated in any frame. In the BF, the only non-zero component is
\begin{equation}
  \langle P_0\rangle\;=\;2\bar\gamma F_1.
\label{z:mean}\end{equation}
In an arbitrary frame, other non-zero components exist. The BF is
a good frame in which to compare the $T=0$ and $T>0$ results because
of the simplicity of the vacuum results in this frame.

The variance of the radiated momentum has the tensor decomposition
\begin{equation}
   V^0_{\mu\nu}\;=\;\int \tilde{dk}
      J_\lambda(k)J^\lambda(k) k_\mu k_\nu\;=\;
    \sum_{i=1}^4 F^0_i {\cal Z}^i_{\mu\nu},
\label{var:decompose}\end{equation}
where the four Lorentz tensors are listed in
eq.~(\ref{ap:z:tensors}). The coefficient functions and tensor
components can be computed as explained in Appendix \ref{ap:tensor}.

We need to write the four contractions
\begin{equation} \begin{array}{rll}
  I^0_1\;=\;&{\cal Z}^1_{\mu\nu}V^0_{\mu\nu}
     &\;=\;0,\\
  I^0_2\;=\;&{\cal Z}^2_{\mu\nu}V^0_{\mu\nu}
     &\;=\;{\alpha\over2\pi} \eta^2 Q^2
       \left[2\Psi\Omega_4-\Omega_0-\Omega_5\right],\\
  I^0_3\;=\;&{\cal Z}^3_{\mu\nu}V^0_{\mu\nu}
     &\;=\;{\alpha\over2\pi} \eta^2 Q^2
       \left[2\Psi\Omega_4-\Omega_0-\Omega_5\right],\\
  I^0_4\;=\;&{\cal Z}^4_{\mu\nu}V^0_{\mu\nu}
     &\;=\;{\alpha\over2\pi} \eta^2 Q^2
       \left[2\Omega_0-{2\over\Psi}\Omega_4\right].
\end{array} \label{var:integrals}\end{equation}
The final expression for each contraction is written in the BF.
The coefficient functions are obtained by solving a linear
system of equations, as explained in Appendix \ref{ap:tensor}.

Since the matrix ${\bf Z}$ (${\bf Z}_{ij}={\cal Z}^i_{\mu\nu}
{\cal Z}^j_{\mu\nu}$, see eq.~\ref{ap:z:dots})
is symmetric under interchange of
${\cal Z}^2$ and ${\cal Z}^3$, and $I^0_2=I^0_3$, one must
have $F^0_2=F^0_3$. In the hard scattering limit, we find
\begin{equation} \begin{array}{rl}
   F^0_1\;=\;&-{\displaystyle\alpha\over\displaystyle\pi}\eta Q^2,\\
   F^0_2\;=\;&F^0_3\;=\;{\displaystyle\alpha\over\displaystyle\pi}
        \eta Q^2{\bar\gamma}^{-2} \left[
          \log\left({\displaystyle Q\over\displaystyle  m}\right)
                        -1\right],\\
   F^0_4\;=\;&{4\displaystyle\alpha\over\displaystyle\pi}\eta Q^2.
\end{array} \label{var:forms}\end{equation}
Note that $F_2=F_3$, as expected, and that collinear
singularities appear in these two coefficient functions.

The tensor components of the covariance can now be written
down. We find that in the BF only the diagonal components
survive.
\begin{equation} \begin{array}{rl}
   V^0_{00}\;=\;&{\displaystyle\alpha\over\displaystyle\pi}\eta Q^2
       \left[2\log\left({\displaystyle Q\over\displaystyle m}\right)
                -1\right],\\
   V^0_{11}\;=\;&V^0_{22}\;=\;{\displaystyle\alpha\over\displaystyle\pi}
       \eta Q^2,\\
   V^0_{33}\;=\;&{\displaystyle\alpha\over\displaystyle\pi}\eta Q^2
       \left[2\log\left({\displaystyle Q\over\displaystyle m}\right)
                -3\right].
\end{array} \label{var:tensors}\end{equation}
In a general frame, of course, there will be other components
of the covariance. The coefficient functions,
eq.~(\ref{var:forms}), however, are Lorentz invariant, and
may be extracted from measurements in any frame.

Other important details to notice about these expressions are
that there are no off-diagonal elements, {\sl i.e.\/}, there
are no angular correlations. Also, the transverse asymmetry
$V^0_{11}-V^0_{22}$ vanishes, and neither of the transverse
components of the variance involves any large logarithms.
Finally, the logarithms in the other two components of the
variance are equal, so that the longitudinal asymmetry
$V^0_{00}-V^0_{33}$ is free from these terms.

\section{Finite Temperature\label{ft}}

The finite temperature piece of the resummed distribution is
even in $P$. As a result, the $T>0$ contribution to all odd
cumulants vanish. In this section we consider only the first
non-vanishing cumulant, {\sl i.e.\/}, the covariance. It has
the tensor decomposition
\begin{equation}
   V_{\mu\nu}^{\scriptscriptstyle T} \;=\;\int \tilde{dk}
      J_\lambda(k)J^\lambda(k) 2B(k\cdot u;T) k_\mu k_\nu\;=\;
    \sum_{i=1}^7 F^{\scriptscriptstyle T}_i {\cal T}^i_{\mu\nu},
\label{ft:decompose}\end{equation}
where our choice of the seven Lorentz tensors is listed in
eq.~(\ref{ap:t:tensors}). Recall that at $T=0$ only four tensors
could be constructed. The three extra tensors are due to the extra
vector $u$ in the problem. The coefficient functions and tensor
components can be computed as before.

We begin by evaluating the contractions $I^{\scriptscriptstyle T}_i=
{\cal T}^i_{\mu\nu}V^{\scriptscriptstyle T}_{\mu\nu}$. These
integrals are most conveniently performed in the rest frame of the
heat bath (HBF). As discussed in
Appendix \ref{ap:ints}, when $Q\gg T$ the contractions are given to
accuracy of order $\exp(-Q/T)$ by
\begin{equation} \begin{array}{rl}
  I^{\scriptscriptstyle T}_1\;=\;&
     -I^{\scriptscriptstyle T}_5\;=\;
    {\displaystyle \pi\alpha\over\displaystyle 3}T^2
        \left[\Omega_2+\Omega'_2-2\Psi\Omega_3\right],\\
  I^{\scriptscriptstyle T}_2\;=\;&
    {\displaystyle \pi\alpha\over\displaystyle 3}T^2
        \left[2\gamma(\Omega_1+\Omega'_6-2\Psi\Omega'_1)
           -(\Omega_0+\Omega'_5-2\Psi\Omega'_4)\right]
       +\gamma^2I^{\scriptscriptstyle T}_5,\\
  I^{\scriptscriptstyle T}_3\;=\;&
    {\displaystyle \pi\alpha\over\displaystyle 3}T^2
        \left[2\gamma'(\Omega'_1+\Omega_6-2\Psi\Omega_1)
           -(\Omega_0+\Omega_5-2\Psi\Omega_4)\right]
       +\gamma'^2I^{\scriptscriptstyle T}_5,\\
  I^{\scriptscriptstyle T}_4\;=\;&
    {\displaystyle \pi\alpha\over\displaystyle 3}T^2
        \biggl[2\Omega_0
           +\left({\gamma'\over\Psi}-2\gamma\right)\Omega_1
           +\left({\gamma\over\Psi}-2\gamma'\right)\Omega'_1\\
                   &\quad\qquad
        -{1\over\Psi}(\Omega_4+\Omega'_4-\gamma\Omega_6
           -\gamma'\Omega'_6)\biggr]
          +{\gamma\gamma'\over\Psi}I^{\scriptscriptstyle T}_5,\\
  I^{\scriptscriptstyle T}_6\;=\;&
    {\displaystyle \pi\alpha\over\displaystyle 3}T^2
        \left[2\Psi\Omega'_1-\Omega_1-\Omega'_6\right]
       -I^{\scriptscriptstyle T}_5,\\
  I^{\scriptscriptstyle T}_7\;=\;&
    {\displaystyle \pi\alpha\over\displaystyle 3}T^2
      {1\over\gamma'}
        \left[2\Psi\Omega_1-\Omega'_1-\Omega_6\right]
       -I^{\scriptscriptstyle T}_5.
\end{array} \label{ft:integs}\end{equation}
We have used the approximations given in eqs.~(\ref{ap:approx}) and
(\ref{ap:halfisfull}). Note that the two pairs
$I^{\scriptscriptstyle T}_2$ and $I^{\scriptscriptstyle T}_3$,
and $I^{\scriptscriptstyle T}_6$ and $I^{\scriptscriptstyle T}_7$
are related by interchange of primed and unprimed quantities. The
angular integrals are listed in Table \ref{ap:angtab}.

We write down only the leading behaviour (largest power of $\bar\gamma$)
of the coefficient functions in the hard scattering limit---
\begin{equation}\begin{array}{rl}
F^{\scriptscriptstyle T}_1\;=\;&
  -{\displaystyle\pi\alpha T^2\over\displaystyle3}
   {\displaystyle 2\log(1+A)\over\displaystyle A},\\
F^{\scriptscriptstyle T}_2\;=\;&
  {\displaystyle\pi\alpha T^2\over\displaystyle3}
       {\displaystyle1\over\displaystyle\bar\gamma^2}
    \,{\displaystyle1\over\displaystyle\gamma_-^2A^2}
   \biggl[A(1-2A)+2A^2\log(2\bar\gamma)\\&\qquad\qquad\qquad\qquad
     -(1+2A-A^2)\log\gamma_--(1+A)^2\log\gamma_+\biggr],\\
F^{\scriptscriptstyle T}_3\;=\;&
  {\displaystyle\pi\alpha T^2\over\displaystyle3}
       {\displaystyle1\over\displaystyle\bar\gamma^2}
    \,{\displaystyle1\over\displaystyle\gamma_+^2A^2}
   \biggl[A(1-2A)+2A^2\log(2\bar\gamma)\\&\qquad\qquad\qquad\qquad
     -(1+A)^2\log\gamma_--(1+2A-A^2)\log\gamma_+\biggr],\\
F^{\scriptscriptstyle T}_4\;=\;&
  {\displaystyle\pi\alpha T^2\over\displaystyle3}
   {\displaystyle 4[A-\log(1+A)]
    \over\displaystyle A^2},\\
F^{\scriptscriptstyle T}_5\;=\;&
  {\displaystyle\pi\alpha T^2\over\displaystyle3}
   \left[4\log(2\bar\gamma)-2\right],\\
F^{\scriptscriptstyle T}_6\;=\;&
  {\displaystyle\pi\alpha T^2\over\displaystyle3}
   {\displaystyle2\over\displaystyle A}
       [2A\log(2\bar\gamma)-A-(1-A)\log\gamma_--(1+A)\log\gamma_+],\\
F^{\scriptscriptstyle T}_7\;=\;&
  {\displaystyle\pi\alpha T^2\over\displaystyle3}
   {\displaystyle2\over\displaystyle A}
        [2A\log(2\bar\gamma)-A-(1+A)\log\gamma_--(1-A)\log\gamma_+].\\
\end{array} \label{ft:forms}\end{equation}
In order to display these results in a compact form, we have introduced
the notation
\begin{equation}
x=\beta_u\cos\chi,\quad\gamma_\pm=\gamma_u(1\pm x),\quad
A=\gamma_+\gamma_--1=\gamma_u^2\beta_u^2\sin^2\chi,
\label{ft:mnemonics}\end{equation}
where $\beta_u$, $\gamma_u$ and $\chi$ are defined in eq.~(\ref{ap:bf}).
Since we are interested in the hard-scattering limit, we have only
kept the leading power of $\bar\gamma$ in these expressions. Note the
appearance of $\log(2\bar\gamma)=\log(Q/m)$ in several of the coefficient
functions.

Although the integrals in eq.~(\ref{ft:integs}) were performed in
the HBF, the coefficient functions are invariant. However, the
components of the covariance tensor are frame dependent. For ease
of comparison with the $T=0$ results in eq.~(\ref{var:tensors}),
we write the non-vanishing components in the BF---
\begin{equation}\begin{array}{rl}
 V^{\scriptscriptstyle T}_{00}\;=\;&
   \left({\displaystyle\pi\alpha T^2\over\displaystyle3}\right)
   {\displaystyle2\over\displaystyle(1+A)(1-x^2)}
    \biggl[(1+x^2)\log\left({\displaystyle Q^2\over\displaystyle m^2}\right)
       +(2A-1-3x^2)\\&\qquad\qquad\qquad\qquad\qquad
          +(1+x)^2\log\gamma_-+(1-x)^2\log\gamma_+\biggr],\\
 V^{\scriptscriptstyle T}_{11}\;=\;&
   \left({\displaystyle\pi\alpha T^2\over\displaystyle3}\right)
   2\left[2-{\displaystyle1\over\displaystyle A}\log(1+A)\right],\\
 V^{\scriptscriptstyle T}_{22}\;=\;&
   \left({\displaystyle\pi\alpha T^2\over\displaystyle3}\right)
   {\displaystyle2\over\displaystyle A}\log(1+A),\\
 V^{\scriptscriptstyle T}_{33}\;=\;&
   \left({\displaystyle\pi\alpha T^2\over\displaystyle3}\right)
   {\displaystyle2\over\displaystyle(1+A)(1-x^2)}
    \biggl[(1+x^2)\log\left({\displaystyle Q^2\over\displaystyle m^2}\right)
       +(2Ax^2-x^2-3)\\&\qquad\qquad\qquad\qquad\qquad
         +(1+x)^2\log\gamma_-+(1-x)^2\log\gamma_+\biggr],\\
 V^{\scriptscriptstyle T}_{01}\;=\;&
   \left({\displaystyle\pi\alpha T^2\over\displaystyle3}\right)
   {\displaystyle4\over\displaystyle(1-x^2)}\beta_u\sin\chi,\\
 V^{\scriptscriptstyle T}_{03}\;=\;&
   \left({\displaystyle\pi\alpha T^2\over\displaystyle3}\right)
   {\displaystyle2\over\displaystyle(1+A)(1-x^2)}
    \biggl[2x\log\left({\displaystyle Q^2\over\displaystyle m^2}\right)
       +2x(A-2)+\\&\qquad\qquad\qquad\qquad\qquad
          (1+x)^2\log\gamma_--(1-x)^2\log\gamma_+\biggr],\\
 V^{\scriptscriptstyle T}_{13}\;=\;&
   \left({\displaystyle\pi\alpha T^2\over\displaystyle3}\right)
   {\displaystyle4\over\displaystyle(1-x^2)}x\beta_u\sin\chi.
\end{array}\label{ft:covform}\end{equation}
A check is provided by the fact that the diagonal elements of this
tensor are positive.

We find several interesting features in eq.~(\ref{ft:covform}). Most
obviously, there are some non-zero off-diagonal components of the
covariance tensor. Recall that $p$, $p'$ and $u$ together define
the reaction plane. Then all in-plane components of $P$ are correlated
with each other, but not with the off-plane component. Furthermore the
transverse components of $P$ no longer have equal variance. This is
also due to the distinction between in-plane and off-plane components.
Finally, all the logarithmically divergent terms arise only in the energy
and longitudinal parts of the tensor. Exactly as at $T=0$, the
coefficients of the logarithms in the diagonal components are equal.

An invariant comparison of the vacuum ($T=0$) and $T>0$ results can be made
using the coefficient functions. For this we need to convert the tensor
expansion in eq.~(\ref{ft:decompose}) into an expansion in terms of
the set $S^i_{\mu\nu}$ (eq.~\ref{ap:t:vartensors}), since the first
four tensors in this set are precisely equal to the tensor set used
at $T=0$. The coefficient functions in this basis are $\tilde
F^{\scriptscriptstyle T}_i=\sum_j F^{\scriptscriptstyle T}_j
{\bf S}^{ij}$. Using the explicit form for ${\bf S}$ given in
eq.~(\ref{ap:t:transf}), we see that $\tilde F^{\scriptscriptstyle T}_i
=F^{\scriptscriptstyle T}_i$ for $i=1,\cdots,4$. A comparison of
eq.~(\ref{var:forms}) with the first four expressions in
eq.~(\ref{ft:forms}) shows the change due to the temperature.
Apart from the differences in the coefficients of the $\log(Q/m)$
terms, there are new $\log\gamma_\pm$ terms. Using the arguments in
Appendix \ref{ap:ints} (see the discussion following
eq.~\ref{ap:halfisfull}) it is easy to see that these are terms in
$\log(Q/T)$.

\begin{figure}
\vskip6truecm
\includegraphics{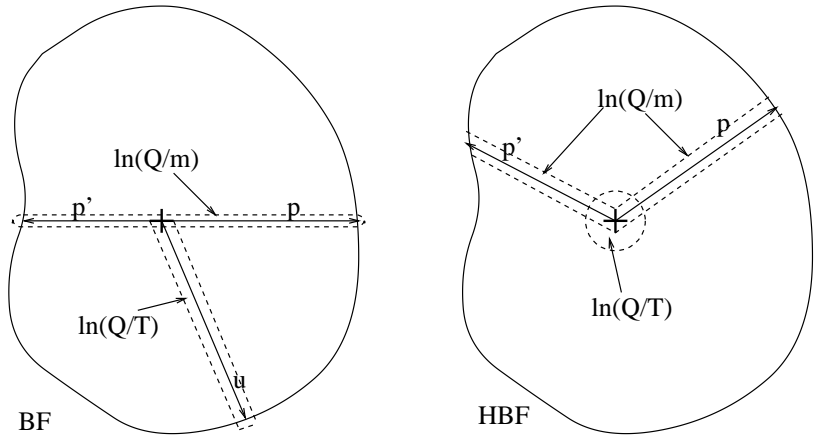}
\caption[dummy]{The regions of the photon phase space from which
   one obtains different logarithmic terms are included within
   dashed lines. The directions of the vectors in the problem are
   shown. Note that in the HBF the origin (marked with a cross) is
   not included in the integration because of infra-red cancellations.
   In the BF the ray parallel to $u$ is removed.}
\label{fg:phsp}\end{figure}

The origin of the logarithms is shown in Figure \ref{fg:phsp}.
The $\log(Q/m)$ terms arise from the same source at $T=0$ and $T>0$.
They come from the part of phase space where the photon momentum $k$
is nearly parallel to $p$ or $p'$. The $\log(Q/T)$ terms, on the
other hand, arise entirely from the thermal distribution. This is
clear from eq.~(\ref{ft:decompose}), since the $\gamma_u$ dependence
of the results can come only from the $k\cdot u$ term when the integrals
are performed in the BF. Hence, these logarithms can be identified as due
to the part of the phase space where $k\cdot u\ll T$. Since this is a
Lorentz invariant 0, we can transform this to the HBF, where it
becomes equivalent to saying that the logarithms arise in the region of
phase space where $k_0\ll T$.

The $\log(Q/T)$ terms never give large contributions. One way of
understanding this is to note that the actual singularity in the Bose
distribution is removed through a cancellation between the real and
virtual diagrams. This is equivalent to
removing the origin in the phase space of $k$. Thus, in the limit
$T\to0$, when the Bose distribution becomes concentrated on the origin,
its contribution to the process actually vanishes. In other words,
positive powers of $T$ always regulate the divergence in $\log T$.
Contrast this to the collinear parts. Along the rays parallel to $p$
or $p'$ in phase space there are actual singularities. These regions
become available as $m\to0$, and hence give rise to genuinely large
contributions to various observables.

When $Q>T\gg m$, then the covariances provide straightforward
measurements of the temperature. Useful observables are the two
asymmetries
\begin{equation}
  A_L\;=\;{\displaystyle\langle P_0P_0\rangle_c-\langle P_3P_3\rangle_c
    \over\displaystyle\langle P_0P_0\rangle_c+\langle P_3P_3\rangle_c},
  \quad\quad
  A_T\;=\;{\displaystyle\langle P_1P_1\rangle_c-\langle P_2P_2\rangle_c
    \over\displaystyle\langle P_1P_1\rangle_c+\langle P_2P_2\rangle_c},
\label{observables:a}\end{equation}
and the correlation functions
\begin{equation}
  C_{\mu\nu}\;=\;{\displaystyle\langle P_\mu P_\nu\rangle_c\over
    \displaystyle\sqrt{\langle P_\mu P_\mu\rangle_c
                         +\langle P_\nu P_\nu\rangle_c}}.
\label{observables:c}\end{equation}
In the BF, $A_T$, $C_{01}$, $C_{03}$ and $C_{13}$ are precisely zero at
$T=0$, but become non-zero at finite temperatures. Each of these is
proportional to $(T/\eta Q)^2$ and hence need not be very small. As a
result, they can yield measurements of the temperature. Furthermore,
all other correlations remain zero even at finite temperature. The
pattern of vanishing and non-vanishing correlations, or equivalently,
off-diagonal components of the covariance tensor shows whether or not
any observed effects are due to finite temperatures. As a result,
these observations provide not only measurements of the temperature,
but also cross checks that the effects do indeed come from thermal effects.

We do not discuss numbers here since such estimates require matching of
the perturbative and resummed parts of the cross section \cite{matching}.
It would nevertheless be interesting to carry out such a computation to
see whether it is possible to design special purpose experiments to
measure these effects.

\section{Discussion and Conclusions\label{results}}

We have performed a resummation of soft photon emissions in two
hard processes in finite temperature QED. These are the scattering
of an electron by a hard spacelike photon, and the annihilation of
an electron positron pair into a hard timelike photon.

The summation has been performed in a fully Lorentz-invariant
manner. This was achieved by separating soft and hard photons
invariantly, eq.~(\ref{mode}), and by writing the thermal photon
propagator invariantly, eq.~(\ref{prop}). This second step
reveals that finite temperature physics involves one extra
Lorentz vector over the corresponding vacuum process \cite{wel82}.
This is a Lorentz-invariant way of stating that a heat-bath selects
out a preferred frame.

We showed that an invariant resummation of soft photon effects is
possible and yields a perfectly finite distribution of the total
radiated photon momentum, $P$. Thus Weldon's extension \cite{wel94}
of the Bloch-Nordsieck theorem \cite{bn,jr} can itself be extended
from energy distributions to the full 4-momentum distribution. The
net effect of finite temperatures is to produce correlations in the
radiation pattern. By performing a tensor decomposition of the
cumulants of $P$, we obtained a Lorentz-invariant comparison of the
radiation pattern in vacuum and in a heat-bath in terms of certain
scalar coefficient functions, eqs.~(\ref{var:forms}) and
(\ref{ft:forms}). At finite temperature, large logarithms appear
in unfamiliar places.

The comparison of these results is simplest in the BF. In this
frame there is a rotational symmetry around the common direction of
the two fermion momenta at $T=0$. Furthermore, there is a symmetry
in the interchange of the two momenta. As a result, the mean momentum
vanishes, leaving only a non-zero mean radiated energy,
eq.~(\ref{z:mean}). Also, there are no correlations between the
different components of $P$, {\sl i.e.\/}, the covariance tensor is
diagonal. The transverse diagonal components of this tensor are
equal due to the rotational symmetry. Large logarithms show up in
the other two diagonal components, eq.~(\ref{var:tensors}).

At finite temperatures both these symmetries are spoilt due to the
presence of the extra vector $u$. The mean of $P$ is not changed,
since momentum can be absorbed from the heat bath as well as emitted.
For a process in equilibrium, these two rates balance out, as
seen from the symmetry of $R_{\scriptscriptstyle T}$
(eq.~\ref{expon}) under $x\to-x$. However, the covariance tensor
changes substantially (eq.~\ref{ft:covform}). The extra
vector, $u$, allows an invariant distinction between transverse
momentum in the reaction plane and off it. The variances of these two
components are no longer related by any symmetry, and hence can be
different. All the in-plane components of $P$ are correlated with
each other, but not with the off-plane component.
As a result, the covariance tensor is no longer diagonal.
In particular, the covariance of the energy and longitudinal momenta
contains a large logarithm, and may be easily measured.

We notice that the $T>0$ results contain two types of logarithms---
either in $Q/m$ or in $Q/T$. The source of the former is the same in
the vacuum and finite temperature theories; they arise from parts
of the phase space where the radiated (or absorbed) photon is
nearly collinear with the fermion momenta. In the $m\to0$ limit
photons can become precisely collinear with the fermions. This
configuration gives a truly divergent contribution to various
observables. As a result, the $\log(Q/m)$ terms can become large.
The $\log(Q/T)$ terms arise only at $T>0$, and are due to soft parts
of the phase space (in the HBF). The Born-level singularity at $k=0$
is removed by cancellation between real and virtual diagrams, and
these logarithms never give large contributions.
It appears \cite{altherretal} that for hard scattering of thermal
particles the $\log(Q/m)$ terms do not arise. In this sense,
the scattering of external particles in a heat bath seems to be
closer to the familiar situation at zero temperature.

When including hard thermal loops, the formal considerations of
section \ref{s:resum} change in two ways (see Appendix \ref{ap:resum}).
First, the phase space element in eq.~(\ref{dlips}) has to be
modified, by taking into account the absorptive part of the
Braaten-Pisarski resummed photon propagator \cite{htl}. Second,
the current in eq.~(\ref{current}) must be rederived using this
resummed photon propagator in the leading perturbative correction.
The formal results are otherwise unchanged provided the cancellation
of infra-red singularities can be shown.

Consequently, we expect the arguments in the rest of this paper to
remain valid after Braaten-Pisarski resummation. In particular, odd
cumulants will still vanish in the thermal part of the resummed
distribution. There will still be seven coefficient functions for
the covariance tensor, and the zero and non-zero components can still
be identified by the methods developed here. Detailed computations
including hard thermal loops are in progress, and will be reported later.

The extension of these results to one-loop order QCD resummations
is straightforward. We refer to the resummation procedure discussed
in \cite{efs} in the context of gluon radiative corrections to the
Drell-Yan process. The phase space measure there is written in the
Sudakov parametrisation. Nevertheless, it is easy to check that
the resummation of the leading soft singularity in one-loop order
corrections gives an exponentiation of $J_\mu(k)J^\mu(k)$. This
current is very similiar to that in eq.~(\ref{current}). The only
differences are that $m^2=0$ in \cite{efs} and that factors of
$\alpha$ are replaced by $\alpha_s C_F$. Since no non-Abelian
couplings appear in QCD to this order, all of Appendix
\ref{ap:resum} can be translated to that context with only these
changes.

The upshot is that the resummation procedure outlined here also
generalises to one-loop order resummation in finite temperature QCD.
All our formul\ae{} can be carried over by the simple replacement
$\alpha\to\alpha_s C_F$. We expect that the logarithmic mass
singularities can be absorbed into the structure functions of the
external hadrons which undergo hard scattering inside the heat-bath.

There are obvious phenomenological consequences of the computations
presented here. Each off-diagonal component of the covariance tensor
can be measured independently. A separate measurement of the temperature
is obtained from each non-zero component. These estimates must agree
with each other, and from that obtained by measuring
$A_{\scriptscriptstyle L}$ and $A_{\scriptscriptstyle T}$.
A further experimental check is
that all off-plane components of this tensor are indeed zero. These
conditions being fulfilled would verify that the observations are
due to thermal effects, and simultaneously give a measurement of the
temperature. In summary, soft photon resummations yield a new
thermometer in high energy physics.

{\bf Acknowledgements:}\ This work was started during the
``Micro-workshop on Hard Probes of the Quark Gluon Plasma'', held
in TIFR, Bombay, in December 1994. We would like to thank Avijit
Ganguly and Rohini Godbole for useful discussions during the workshop.
We would also like to thank Rohini Godbole for a careful reading of
the manuscript.

%
%
\appendix

\section{The Resummation Procedure\label{ap:resum}}

Here we give a diagrammatic proof of the resummed result to
order $\alpha$ for the processes considered in the text using
the techniques of \cite{lt}. The hard scattering cross sections
with soft thermal photon contributions can be obtained in
perturbation theory. The hard particles are external and do not
thermalise in the heat bath. The interaction Lagrangian is
\begin{equation}
  {\cal L}_I={\cal L}^h_I+{\cal L}^s_I,
\label{ap1:eq1}\end{equation}
where ${\cal L}_I^h$ is written in terms of the hard part of the
photon field and ${\cal L}_I^s$ in terms of the soft part---
\begin{equation}
  {\cal L}^h_I=j^h_\mu(x)\cdot A^\mu_h(x)\qquad{\rm and}\qquad
  {\cal L}^s_I=j^h_\mu(x)\cdot A^\mu_s(x).
\label{ap1:eq2}\end{equation}
In this appendix we use the notation $j^\mu\cdot A_\mu=\int d^4x
j^\mu(x) A_\mu(x)$ and the current $j_\mu=ie\overline\psi\gamma_\mu
\psi$. In order to make the computation transparent, we quantise the
fields in a box of volume V and take infinite volume limit at the end.
Thus,
\begin{equation}
  A^s_\mu(x)=\sum_l'\frac{1}{\sqrt{2 k_l V}}
      [a_l\epsilon_\mu(k_l)\exp(-ik_l\cdot x)+
       a_l^\dagger\epsilon^*_\mu(k_l)\exp(ik_l\cdot x)].
\label{ap1:eq3}\end{equation}
The prime over the summation denotes that it involves only soft modes.
The continuum Lorentz invariant version of this is given in
eq.~(\ref{mode}).

In terms of the transition matrix, ${\bf T}$, and the density matrix
for the thermal fields, $\rho$, the inclusive cross section is
\begin{equation}\begin{array}{rl}
  \sigma\;=\;&{\displaystyle1\over\displaystyle F}
                       Tr (\rho{\bf T}{\bf T}^\dagger)\\
         \;=\;&{\displaystyle1\over\displaystyle Z_0F}
      \displaystyle\sum_{\{m\},\{n\},e'} \exp(-\sum_l\beta n_l\epsilon_l)
         \biggl|\langle\{m\},e'|{\bf T}|\{n\},e,\gamma^*\rangle_c
                \biggr|^2,\\
  {\rm where\ }&Z_0=\prod_l[1-\exp(-\beta\epsilon_l)]^{-1}
      {\rm\ \ and\ \ }i{\bf T}=\exp[i\int dx{\cal L}_I(x)]-1.
\end{array}\label{ap1:eq4}\end{equation}
The subscript $c$ in the matrix element denotes the connected part.
The label $\{n\}$ stands for the set of occupation numbers for each
mode of the soft fields, and $F$ is a flux factor. Note, $\rho=\exp
(-\beta H)/Z$ where $Z$ is the full partition function of the
interacting system. The process of taking connected parts in the
second line absorbs the interaction effects and leaves only a
normalisation by the ideal gas partition function $Z_0$.
The density matrix for
electrons in the heat bath is disregarded since the hard particles
are not thermal averaged and because there are no electron loops to
lowest order.

The soft photon contribution to the hard scattering cross section
can be obtained from the following part of the transition matrix---
\begin{equation}
  {\bf T}\approx T\left[j^\mu\cdot A_\mu^h
      \left(1+j^\mu\cdot A_\mu^s+\frac{1}{2}
             (j^\mu\cdot A_\mu^s)^2+\cdots\right)\right],
\label{ap1:eq5}\end{equation}
where $T$ denotes time ordering. This expansion is made keeping the
order $\alpha$ term in the hard scattering and all orders in the soft
thermal photons. After Wick contractions, this gives two types of
graphs--- real soft photon emission and absorption, and virtual soft
photon corrections. For every field which is thermal averaged the
contraction gives a thermal propagator.

\begin{figure}
\vskip7.5truecm
\includegraphics{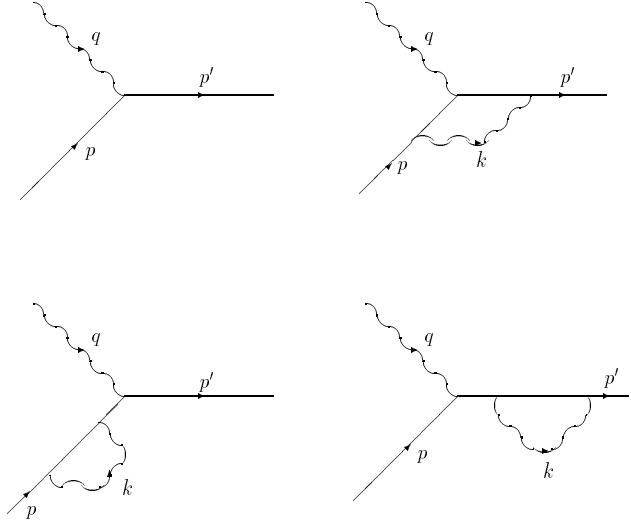}
\caption[dummy]{The virtual corrections discussed in eq.~(\ref{ap1:eq7}).
   The first diagram corresponds to the lowest order process in $\sigma_v$,
   and the other three to the parts in $M_v$.}
\label{fg:feynv}\end{figure}

To order $\alpha$ the virtual soft photon contribution (Figure
\ref{fg:feynv}) to the hard scattering cross section can be obtained
from
\begin{equation}
  {\bf T}_v=\frac{1}{2}T(j^\mu\cdot A_\mu^h\;j^\alpha\cdot
      A_\alpha^s\;j^\beta\cdot A_\beta^s),
\label{ap1:eq6}\end{equation}
by Wick contracting only the soft thermal photons. To this order
there are no thermal fermion loops. Since there are no external
soft photons in the virtual part, only the diagonal elements of
the transition matrix will contribute. Taking the continuum limit,
we obtain the result in an arbitrary frame as
\begin{equation}\begin{array}{rl}
  \sigma_v&\;=\;{\displaystyle e^2\over\displaystyle F}
         \displaystyle\int{\displaystyle d^4p'd^4k
                        \over\displaystyle(2\pi)^8}
      2\pi\delta(p'^2-m^2)\Theta(p'_0)(2\pi)^4 \delta^4(p+q-p')\\
      &\qquad\qquad\qquad\qquad\times
         \bar u(p')\epsilon\!\!\slash(q)u(p)]^*[\bar u(p')M_vu(p)],\\
  {\rm where\ }&M_v=\epsilon_\mu\Gamma^\mu(k,p,p')+\Sigma(k,p')
     {\displaystyle1\over\displaystyle p'\!\!\!\!\slash-m}
                \epsilon\!\!\slash(q)+
     \epsilon\!\!\slash(q){\displaystyle1\over\displaystyle
                p\!\!\!\slash-m}\Sigma(k,p).
\end{array}\label{ap1:eq7}\end{equation}
Here $\Sigma(k,p)$ and $\Gamma_\mu(k,p,p')$ are the usual
self-energy and vertex functions calculated at $T=0$, only with the
photon propagator replaced by the thermal photon propagator of
eq.~(\ref{prop}). The electron propagators do not have thermal parts,
since they are assumed external. In short, this expression
is identical to the $T=0$ result but for the thermal photon
propagator. This remains true even if one considers the Braaten-Pisarski
resummed hot thermal loops \cite{htl}, although the form of the propagator
then changes.

\begin{figure}
\vskip9truecm
\includegraphics{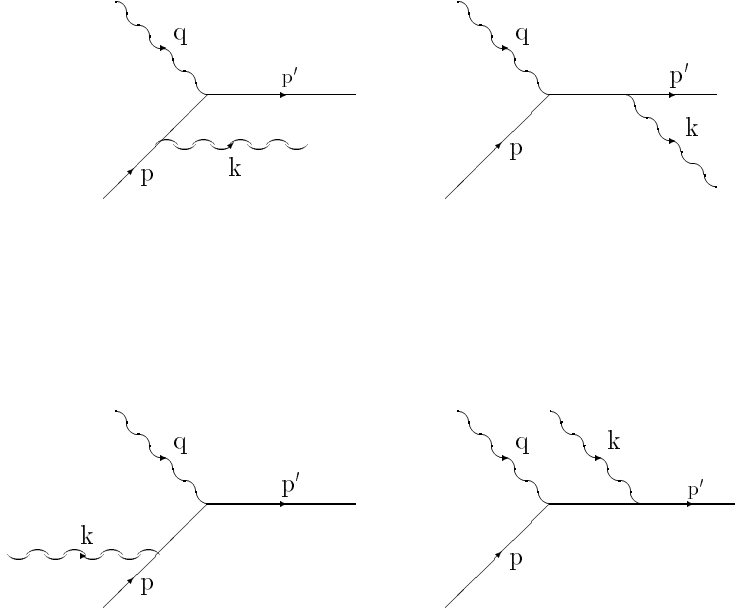}
\caption[dummy]{The real photon emission and absorption corrections
   of $\sigma_r$ in eq.~(\ref{ap1:eq11}). Absorbed soft photons come
   in from the left, emitted photons exit to the right. Note that the
   distinction between emission and absorption is included into the
   phase space measure.}
\label{fg:feynr}\end{figure}

The real soft photon emission and absorption graphs (Figure
\ref{fg:feynr}) upto the same order are obtained from the term
\begin{equation}
{\bf T}_r= T(j^\mu\cdot A_\mu^h\;j^\alpha\cdot A_\alpha^s).
\label{ap1:eq8}\end{equation}
The matrix element of this operator is proportional to
$\langle\{m\}|A^s_\alpha|\{n\}\rangle$.  Only the off-diagonal
elements of the transition matrix contribute to this cross section.
Using the identity
\begin{equation}
\frac{1}{Z_l}\sum_{n_l}n_l\exp(-\beta n_lk_l)=B(k_l;\beta),
\label{ap1:eq9}\end{equation}
(where $\beta=1/T$)
we get the following cross section for real photon emission and
absorption in the 3 limit---
\begin{equation}\begin{array}{rl}
  \sigma_r&\;=\;{\displaystyle e^4\over\displaystyle F}
      \displaystyle\int{\displaystyle d^4p'\over\displaystyle (2\pi)^4}
     2\pi\delta(p'^2-m^2)\Theta(p'_0)\;d\phi(k)\\
    &\quad\qquad\times(2\pi)^4\delta^4(p+q-p'-k)
               \;\biggl|\bar u(p')M_ru(p)\biggr|^2,\\
{\rm where\ }&
   M_r=\epsilon\!\!\slash(q){\displaystyle1\over\displaystyle
             p'\!\!\!\!\slash+k\!\!\!\slash-m}\epsilon^*\!\!\!\!\slash(k)
       +\epsilon^*\!\!\!\!\slash(k){\displaystyle1\over\displaystyle
             p\!\!\slash-k\!\!\!\slash-m}\epsilon\!\!\slash(q),
\end{array}\label{ap1:eq11}\end{equation}
and the phase space element for real thermal photons is given by
eq.~(\ref{dlips})\footnote{In section \ref{s:resum} a Lorentz invariant
cutoff on the soft modes has also been included with this. The rest of
the phase space element was also written down in \cite{wel94}.}.
This phase space for $T>0$ changes because of the additional absorption
processes which do not exist in the vacuum theory. Again, the electron
propagators do not have thermal parts. The important point is that in
both the real and virtual corrections, the matrix elements are the same
as at zero temperature. Finite temperature effects are subsumed into the
thermal photon phase space and propagator.

We note that the squared matrix element appearing in $\sigma_r$ amounts
to taking the one loop corrections to the basic process with a cut on
the soft photon line. The phase space thus comes from the absorptive part
of the thermal photon propagator. Upon performing Braaten-Pisarski
resummation, the real photon phase space element will then be changed in
obvious ways.

The soft photon limit is obtained by setting $k$ to zero in the numerator
of the matrix elements in $\sigma_v$ and $\sigma_r$. In this limit we get
\begin{equation}\begin{array}{rl}
  [\overline u(p')\epsilon\!\!\!\slash u(p)]^*
       &[\overline u(p')M_vu(p)]\;=\;
     -\biggl|\overline u(p')M_ru(p)\biggr|^2 \\
  \;=&\;[\epsilon_\mu(k)J^\mu(k)]^2
      \biggl|\overline u(p')\epsilon\!\!\!\slash(q)u(p)\biggr|^2,
\end{array}\label{ap1:eq12}\end{equation}
where $J^\mu$ is the current in eq.~(\ref{current}). This is the origin
of the factorisation in eq.~(\ref{in:factor}). Taking higher order terms
from eq.~(\ref{ap1:eq5}) gives the exponentiation of the order $\alpha$
result. It is possible to show that a factorisation in the form shown
in eq.~(\ref{ap1:eq12}) holds with higher powers of $(\epsilon_\mu
J^\mu)^2$. As a result, we obtain the squares of the matrix elements
shown in eq.~(\ref{matel}), and hence the generating function of
eq.~(\ref{model}). A full proof will be given elsewhere.

\section{Hard-Scattering Kinematics\label{ap:kinema}}

The problem considered here contains three Lorentz vectors, the
two fermion momenta, $p$ and $p'$, and the velocity vector of the
heat-bath, $u$. With three vectors we can construct three Lorentz
invariants. Since $p^2=p'^2=m^2$ and $u^2=1$, we can work with
the dimensionless invariants
\begin{equation}
  {1\over m} p\cdot u\;=\;\gamma,\quad
  {1\over m} p'\cdot u\;=\;\gamma',\quad{\rm and}\quad
  {1\over m^2} p\cdot p'\;=\;\Psi.
\label{ap:invariants}\end{equation}
For $T=0$, the problem does not involve $u$ and the only invariant
available is $\Psi$.

For the electron scattering problem, it is a convention to use the
square of the vector $q=p'-p$ as one of the invariants. This is a
measure of the scale of the process. We have the relation
\begin{equation}
  Q^2\;\equiv\;-q^2\;=\;2m^2(\Psi-1).
\label{ap:qsqr}\end{equation}
The limit of a hard scattering, $Q^2\to\infty$, corresponds to
taking $\Psi\to\infty$. The momentum $q'=p+p'$ is closely related.
In the annihilation process this corresponds to the final photon
momentum, and its square to its invariant mass. We have,
\begin{equation}
  q'^2\;=\;2m^2(\Psi+1)\stackrel{\Psi\to\infty}{\longrightarrow} Q^2.
\label{ap:qpsqr}\end{equation}
Note that $q$ is space-like and $q'$ time-like, and $q.q'=0$. We
choose to use $q'$ to set the scale of the hard scattering process.

We shall find it most convenient to work in one of two frames,
either the rest frame of the heat-bath (HBF) or the Breit
frame (BF). In this latter frame, $p$ and $p'$ have equal and
opposite 3-momenta, and their common direction conventionally
defines the $z$-direction. For the annihilation process the BF
is just the rest frame of the final off-shell photon. The reaction
plane in the HBF is chosen to be the $xz$-plane. Then this is also
the plane containing $p$, $p'$ and $u$ in the BF.

We find it useful to parametrise different momenta in the BF
as displayed below
\begin{equation} \begin{array}{rl}
  p\;=&\;m\bar\gamma(1,0,0,\bar\beta),\\
  p'\;=&\;m\bar\gamma(1,0,0,-\bar\beta),\\
  u\;=&\;\gamma_u(1,\beta_u\sin\chi,0,\beta_u\cos\chi),\\
  q'\;=&\;2m\bar\gamma(1,0,0,0).
\end{array} \label{ap:bf}\end{equation}
Note that the BF is the rest frame of $q'$.
The parametrisations in the HBF can be obtained by boosting these
so as to obtain $u=(1,0,0,0)$.

The quantities $\bar\gamma$, $\gamma_u$ and $\chi$ are Lorentz
invariant. It is easy to see that
\begin{equation}
  \bar\beta\;=\;\sqrt{{\Psi-1}\over{\Psi+1}},\qquad{\rm and}
  \qquad\bar\gamma\;=\;\sqrt{{\Psi+1}\over2}.
\label{ap:barred}\end{equation}
As a result, the hard-scattering limit corresponds to taking
$\bar\gamma\to\infty$. In addition, we can write an explicitly
Lorentz-invariant form for $\gamma_u$,
\begin{equation}
  \gamma_u\;=\;{\displaystyle u\cdot(p+p')\over
                \displaystyle\sqrt{(p+p')^2}}.
\label{ap:gmu}\end{equation}
An explicitly invariant form of $\chi$ can also be written down.

\section{The Integrals\label{ap:ints}}

In this appendix we investigate integrals of the form
\begin{equation}
  I\;=\;\int {d^4k\over(2\pi)^4} 2\pi\delta(k^2)\Theta(k_0)
    \Theta\left(\eta q'^2-q'\cdot k\right) 2B(k\cdot u;T) F(k),
\label{ap:integraldef}\end{equation}
where $B$ is the Bose distribution function and $F(k)$ is a
scalar involving (possibly) $p$, $p'$ and $u$. The second
$\Theta$-function on the right restricts the phase space to
soft modes. Since we are interested in the region $\eta\ll1$,
we shall freely use the hard scattering kinematics of
Appendix \ref{ap:kinema}.

In the BF this phase space constraint takes the particularly
simple form
\begin{equation}
  \Theta\left(\eta q'^2-q'\cdot k\right)\;=\;
  \Theta\left(2m\eta\bar\gamma-k_0\right).
\label{ap:constraintbf}\end{equation}
The only constraint is on the values of $k_0$.
For $T=0$, the Bose distribution drops out, and the only
possible angular dependences are in $F(k)$. Hence, all the
relevant integrals are best done in the BF.

For $T>0$, the Bose distribution picks up angular factors in the
BF. Hence, it is simpler to perform these integrals in the HBF.
However, the constraint now cuts off part of the angular region
in phase space. Fortunately, the analysis of this problem is not
complicated. The constraint can be written as
\begin{equation}
  \Theta\left(\eta q'^2-q'\cdot k\right)\;=\;
  \Theta\left({2m\eta\bar\gamma\over\gamma_u}-k_0(1-\beta_u
    \hat{\bf q}'\cdot\hat{\bf k})\right),
\label{ap:constrainthbf}\end{equation}
where boldfaced symbols with carets denote unit 3-vectors.

The quantity within brackets on the right hand side lies between
$1-|\beta_u|$ and $1+|\beta_u|$. As a result, the constraint
picks a non-trivial angular region of integration only for
$k_-\le k_0\le k_+$, where
\begin{equation}
  k_-\;=\;2m\eta\bar\gamma\sqrt{{1-|\beta_u|\over1+|\beta_u|}},\quad
   k_+\;=\;2m\eta\bar\gamma\sqrt{{1+|\beta_u|\over1-|\beta_u|}}.
\label{ap:limits}\end{equation}
For $k\le k_-$, the full angular region is allowed, and for
$k\ge k_+$, no phase space volume is allowed.

Now, the integral $I$ can be split up into an integration over
$0\le k_0\le k_-$ ($I_1$) and another over $k_-\le k_0\le k_+$
($I_2$). The angular integral in the first part is over the full
sphere $\hat{\bf k}$, whereas that in the second part is
restricted by the constraint.

This second integral becomes
\begin{equation}
  I_2\;=\;{1\over 8\pi^3}\int_{k_-}^{k_+}kdkB(k;T)f(k),
\label{ap:eye2}\end{equation}
where $f(k)$ is obtained from $F(k)$ by the appropriate
angular integration. Since $B(k;T)$ is monotonically decreasing,
we have the bound
\begin{equation}
  \left|I_2\right|\;\le\; {B(k_-;T)\over8\pi^3}
    \int_{k_-}^{k_+} kdk|f(k)|.
\label{ap:bound}\end{equation}
For all the integrands we are interested in, the growth with
$\bar\gamma$ of the integral on the right hand side is
polynomially bounded. Since $B(k_-;T)$ decreases exponentially
with $\bar\gamma$, $I_2$ is exponentially small in the hard
scattering limit. As a result, it can be neglected in comparison
with $I_1$.

The angular integrals in the first part are unconstrained. Hence
we find it useful to define the integrals
\begin{equation}
  \Omega_i=\int d\Omega F_i(k),\qquad{\rm where}\qquad
    d\Omega={d\cos\theta d\phi\over4\pi},
\label{ap:angular}\end{equation}
with the convention that $F_i(k)$ is rendered dimensionless
by dividing each factor of $p$ or $p'$ by $m$ and $k$ by $k_0$.
In the full computation we need many forms of $F_i(k)$. These,
and the resulting angular integrals are listed in Table
\ref{ap:angtab}.

For $Q\to\infty$, we can now write
\begin{equation}
  I\;\approx\;{\Omega_i\over2\pi^2}\int_0^{k_-}kdkB(k;T)
     F(k,\theta,\phi).
\label{ap:approx}\end{equation}
In actual applications, we shall choose $k_-$ to be large enough
that
\begin{equation}
  \int_0^{k_-} dk k B(k;T)\;\approx\;{\pi^2 T^2\over6}.
\label{ap:halfisfull}\end{equation}
A numerical evaluation of the incomplete Bose integral on the
left shows that this approximation is valid to an accuracy of
one part in thousand for $k_-\approx8.6T$. Now if we were to
have $T\ll Q$, then we must either take $\eta\ll 1$ or
$1-|\beta_u|\ll 1$ in order to satisfy this condition. Since
$\eta$ is some fixed number, for $Q/T\gg1$, it is possible to
satisfy the constraint even for $\gamma_u\gg1$.

In short, in the hard-scattering limit, we can approximate all
integrals of the form in eq.~(\ref{ap:integraldef}) with
exponential accuracy to obtain
\begin{equation}
   I_i\;=\;{\displaystyle T^2\Omega_i\over\displaystyle 12}.
\label{ap:intres}\end{equation}
The angular integrals $\Omega_i$ are frame dependent, and are
listed in Table \ref{ap:angtab}.

\begin{table}
  \begin{tabular}{|c|c|c|c|}  \hline
  Integral & $F_i$ & HBF & BF \\ \hline
  $\Omega_0$ & $1$ & $1$ & $1$ \\ \hline
  $\Omega_1$ & $[m k_0/p\cdot k]$ & $L(\beta)/\gamma$ &
               $L(\bar\beta)/\bar\gamma$ \\ \hline
  $\Omega_2$ & $[m^2 k_0^2/(p\cdot k)^2]$ & $1$ & \\ \hline
  $\Omega_3$ & $[m^2 k_0^2/p\cdot k\,p'\cdot k]$ &
               $L(\beta_{\scriptscriptstyle\Psi})/\Psi$ & \\ \hline
  $\Omega_4$ & $[p'\cdot k/p\cdot k]$ &
               $[\phi+\Phi L(\beta)/\gamma]/(\beta\gamma)^2$ &
               $2L(\bar\beta)-1$ \\ \hline
           & & $[\xi+\lambda L(\beta)]/(\beta\gamma)^4$, &\\
  $\Omega_5$ & $[p'\cdot k/p\cdot k]^2$ &
               $\xi=\Phi^2+2\phi^2-\Gamma^2$, &
               $4\bar\gamma^2+1-4L(\bar\beta)$ \\
           & & $\lambda=\Gamma^2-\phi(\phi-2\Phi/\gamma)$. &\\ \hline
  $\Omega_6$ & $mk_0[p'\cdot k/(p\cdot k)^2]$ &
               $[\Phi+\phi L(\beta)/\gamma]/(\beta\gamma)^2$ &
               $2\bar\gamma-L(\bar\beta)/\bar\gamma$ \\ \hline
  \end{tabular}
  \caption[dummy]{Definitions and values of the angular integrals
     in the HBF and BF. The integration is over the full sphere
     $\hat{\bf k}$.We have used the mnemonics (a)
     $L(x)=\log\bigl((1+x)/(1-x)\bigr)/2x$, (b)
     $\beta_{\scriptscriptstyle\Psi}=\sqrt{1-1/\Psi^2}$, (c)
     $\phi=\gamma\gamma'-\Psi$, (d) $\Phi=\Psi\gamma-\gamma'$
     (e) $\Phi'=\Psi\gamma'-\gamma$ and (f) $\Gamma=\beta\beta'
     \gamma\gamma'$. $\Omega'_i$ are defined
     by interchanging $p$ and $p'$. In the HBF, their values are
     obtained by interchanging primed and unprimed quantities.
     In the BF, $\Omega_i=\Omega'_i$, since each integral is
     symmetric under $\bar\beta\to-\bar\beta$.}
\label{ap:angtab}\end{table}

\section{Tensor Decompositions\label{ap:tensor}}

The cumulants are symmetric Lorentz tensors which can be written
as an integral $I_{\mu_1\mu_2\cdots\mu_n}$ over a vector $k$,
containing some vectors $V_1$, $V_2$, {\sl etc\/}, in the integrand.
As a result, they admit expansions of the form---
\begin{equation}
  I_{\mu_1\mu_2\cdots\mu_n}\;=\;\sum_i
        F_i Q^i_{\mu_1\mu_2\cdots\mu_n}.
\label{ap:formfac}\end{equation}
The set of basis tensors $Q^i$ consists of all possible symmetric
rank-$n$ tensors which may be written using the vectors $V_1$,
$V_2$, {\sl etc\/}, and the metric tensor $g_{\mu\nu}$. We choose
the basis tensors to be dimensionless.

The coefficients $F_i$ are Lorentz scalars, and given by the
solution of the set of linear equations
\begin{equation}
  {\cal I}_i\;=\;\sum_i {\bf Q}_{ij} F_j,\quad{\rm where}\quad
  {\cal I}_i\;=\;Q^i_{\mu_1\cdots\mu_n}
                            I^{\mu_1\cdots\mu_n},
  \;\;{\bf Q}_{ij}\;=\;Q^i_{\mu_1\cdots\mu_n}
                            Q^{j\,\mu_1\cdots\mu_n}.
\label{ap:linsys}\end{equation}
The matrix of contractions of the basis tensors, ${\bf Q}$, is
symmetric. The integrals ${\cal I}_i$ are Lorentz invariant.

The only vectors involved in the problem at $T=0$ are $p$ and $p'$.
The tensors necessary for the mean momentum are then $p/m$ and $p'/m$.
For the variance, we need to consider the four tensors
\begin{equation} \begin{array}{lr}
    {\cal Z}^1_{\mu\nu}=\;\;\;g_{\mu\nu},&
    {\cal Z}^2_{\mu\nu}={\displaystyle p_\mu p_\nu
                           \over\displaystyle m^2},\\
    {\cal Z}^3_{\mu\nu}={\displaystyle p'_\mu p'_\nu
                           \over\displaystyle m^2},&
    {\cal Z}^4_{\mu\nu}={\displaystyle p_\mu p'_\nu+p_\nu p'_\mu
                           \over\displaystyle2m^2\Psi}.
\end{array} \label{ap:z:tensors}\end{equation}
The corresponding matrix of contractions is
\begin{equation}
  {\bf Z}\;=\;
      \left(\begin{array}{cccc}
         4 & 1 & 1 & 1 \\
         1 & 1 & \Psi^2 & 1 \\
         1 & \Psi^2 & 1 & 1 \\
         1 & 1 & 1 & {\displaystyle 1+\Psi^2\over\displaystyle2\Psi^2}
            \end{array}\right).
\label{ap:z:dots}\end{equation}
The linear system to be solved is ${\bf Z} F={\cal I}$.

For the covariance at $T>0$, the basis tensors must be built out
of three vectors, $p$, $p'$ and $u$. It is more convenient to
work in terms of $u$ and the two vectors $\tilde p=p-(p\cdot u)u$
and $\tilde p'=p'-(p'\cdot u)u$. These two vectors are orthogonal
to $u$ \cite{wel82}. There are seven basis tensors, and they can
be chosen to be
\begin{equation} \begin{array}{lr}
    {\cal T}^1_{\mu\nu}=\;\;\;g_{\mu\nu}-u_\mu u_\nu&
    {\cal T}^2_{\mu\nu}={\displaystyle\tilde p_\mu\tilde p_\nu
                             \over\displaystyle p^2}\\
    {\cal T}^3_{\mu\nu}={\displaystyle\tilde p'_\mu\tilde p'_\nu
                             \over\displaystyle p'^2}&
    {\cal T}^4_{\mu\nu}={\displaystyle\tilde p_\mu \tilde p'_\nu +
        \tilde p_\nu \tilde p'_\mu\over\displaystyle2 p\cdot p'}\\
    {\cal T}^5_{\mu\nu}=u_\mu u_\nu&
    {\cal T}^6_{\mu\nu}={\displaystyle u_\mu \tilde p_\nu + u_\nu \tilde p_\mu
                        \over\displaystyle2 u\cdot p}\\
    {\cal T}^7_{\mu\nu}={\displaystyle u_\mu \tilde p'_\nu +
        u_\nu \tilde p'_\mu\over\displaystyle2 u\cdot p'}.&
\end{array} \label{ap:t:tensors}\end{equation}
With this choice of tensors, the matrix of contractions is
\begin{equation}
  {\bf T}\;=\;\left(\begin{array}{ccccccc}
     3 & -\gamma^2\beta^2 & -\gamma'^2\beta'^2 &
         -{\displaystyle\phi\over\displaystyle\Psi} & 0 & 0 & 0 \\
     -\gamma^2\beta^2 & \gamma^4\beta^4 & \phi^2 &
         {\displaystyle\gamma^2\beta^2\phi\over\displaystyle\Psi} &
         0 & 0 & 0\\
     -\gamma'^2\beta'^2 & \phi^2 & \gamma'^4\beta'^4 &
         {\displaystyle\gamma'^2\beta'^2\phi\over\displaystyle\Psi} &
         0 & 0 & 0\\
     -{\displaystyle\phi\over\displaystyle\Psi} &
         {\displaystyle\gamma^2\beta^2\phi\over\displaystyle\Psi} &
         {\displaystyle\gamma'^2\beta'^2\phi\over\displaystyle\Psi} &
         {\displaystyle \Gamma^2+\phi^2\over\displaystyle 2\Psi^2} &
         0 & 0 & 0\\
  0 & 0 & 0 & 0 & 1 & 0 & 0\\
  0 & 0 & 0 & 0 & 0 & -{\displaystyle\beta^2\over\displaystyle2} &
      -{\displaystyle\phi\over\displaystyle2\gamma\gamma'}\\
  0 & 0 & 0 & 0 & 0 &
      -{\displaystyle\phi\over\displaystyle2\gamma\gamma'} &
     -{\displaystyle\beta'^2\over\displaystyle2}
      \end{array}\right).
\label{ap:t:dots}\end{equation}
We have used the notation for $\phi$ and $\Gamma$ as given in the
caption of Table~\ref{ap:angtab}.

For comparison with $T=0$, it is more convenient to express the
results in terms of the tensor set
\begin{equation}\begin{array}{rl}
    S^i_{\mu\nu}={\cal Z}^i_{\mu\nu}\quad{\rm for\ }i=1,\cdots4,&
       S^5_{\mu\nu}=u_\mu u_\nu,\\
    S^6_{\mu\nu}={\displaystyle u_\mu p_\nu + u_\nu p_\mu
                        \over\displaystyle2 u\cdot p},&
    S^7_{\mu\nu}={\displaystyle u_\mu p'_\nu +
        u_\nu p'_\mu\over\displaystyle2 u\cdot p'}.
\end{array} \label{ap:t:vartensors}\end{equation}
These can be expanded in terms of the set ${\cal T}^i$. We have
$T^i_{\mu\nu}={\bf S}^{ij}S^j_{\mu\nu}$, where
\begin{equation}
  {\bf S}\;=\;\left(\begin{array}{ccccccc}
      1 & 0 & 0 & 0 & -1 & 0 & 0 \\
      0 & 1 & 0 & 0 & \gamma^2 & -2\gamma^2 & 0 \\
      0 & 0 & 1 & 0 & \gamma'^2 & 0 & -2\gamma'^2 \\
      0 & 0 & 0 & 1 & {\displaystyle\gamma\gamma'\over\displaystyle\Psi}
                    & -{\displaystyle\gamma\gamma'\over\displaystyle\Psi}
                    & -{\displaystyle\gamma\gamma'\over\displaystyle\Psi} \\
      0 & 0 & 0 & 0 & 1 & 0 & 0 \\
      0 & 0 & 0 & 0 & -1 & 1 & 0 \\
      0 & 0 & 0 & 0 & -1 & 0 & 1
      \end{array}\right).
\label{ap:t:transf}\end{equation}

\end{document}